\documentstyle[twoside,fleqn]{article}
\makeatletter
\def\fileversion{v2.6}
\def\filedate{24 November 1993}

\newdimen\@bls                    % \@b(ase)l(ine)s(kip)
\@bls=\baselineskip               % \@bls ~= \baselineskip for \normalsize
\advance\@bls -1ex                % (fudge term)
\newdimen\@eps                    %
\def\section{\@startsection{section}{1}{\z@}
  {1.5\@bls plus 0.5\@bls}{1\@bls}{\normalsize\bf}}
\def\subsection{\@startsection{subsection}{2}{\z@}
  {1\@bls plus 0.25\@bls}{\@eps}{\normalsize\bf}}
\def\subsubsection{\@startsection{subsubsection}{3}{\z@}
  {1\@bls plus 0.25\@bls}{\@eps}{\normalsize\bf}}
\def\paragraph{\@startsection{paragraph}{4}{\parindent}
  {1\@bls plus 0.25\@bls}{0.5em}{\normalsize\bf}}
\def\subparagraph{\@startsection{subparagraph}{4}{\parindent}
  {1\@bls plus 0.25\@bls}{0.5em}{\normalsize\bf}}

\def\@sect#1#2#3#4#5#6[#7]#8{\ifnum #2>\c@secnumdepth
  \def\@svsec{}\else 
  \refstepcounter{#1}\edef\@svsec{\csname the#1\endcsname.\hskip0.5em}\fi
  \@tempskipa #5\relax
  \ifdim \@tempskipa>\z@
    \begingroup 
      #6\relax
      \@hangfrom{\hskip #3\relax\@svsec}{\interlinepenalty \@M #8\par}%
    \endgroup
    \csname #1mark\endcsname{#7}\addcontentsline
      {toc}{#1}{\ifnum #2>\c@secnumdepth \else
        \protect\numberline{\csname the#1\endcsname}\fi #7}%
  \else
    \def\@svsechd{#6\hskip #3\@svsec #8\csname #1mark\endcsname
      {#7}\addcontentsline{toc}{#1}{\ifnum #2>\c@secnumdepth \else
        \protect\numberline{\csname the#1\endcsname}\fi #7}}%
  \fi \@xsect{#5}}

\long\def\@makefigurecaption#1#2{\vskip 10mm #1. #2\par}

\long\def\@maketablecaption#1#2{\hbox to \hsize{\parbox[t]{\hsize}
  {#1 \\ #2}}\vskip 0.3ex}

\def\fnum@figure{Figure \thefigure}
\def\figure{\let\@makecaption\@makefigurecaption \@float{figure}}
\@namedef{figure*}{\let\@makecaption\@makefigurecaption \@dblfloat{figure}}

\def\table{\let\@makecaption\@maketablecaption \@float{table}}
\@namedef{table*}{\let\@makecaption\@maketablecaption \@dblfloat{table}}

\textfloatsep=\floatsep           % Space between main text and floats
\intextsep=\floatsep              % Space between in-text figures and 
\long\def\@makefntext#1{\parindent 1em\noindent\hbox{${}^{\@thefnmark}$}#1}

\def\maketitle{\begingroup        % Initialize generation of front-matter
    \def\thefootnote{\fnsymbol{footnote}}%
    \newpage \global\@topnum\z@ 
    \@maketitle \@thanks
  \endgroup
  \let\maketitle\relax \let\@maketitle\relax
  \gdef\@thanks{}\let\thanks\relax
  \gdef\@address{}\gdef\@author{}\gdef\@title{}\let\address\relax}

\def\justify@on{\let\\=\@normalcr
  \leftskip\z@ \@rightskip\z@ \rightskip\@rightskip}

\newbox\fm@box                    % Box to capture front-matter in

\def\@maketitle{%                 % Actual formatting of \maketitle
  \global\setbox\fm@box=\vbox\bgroup
    \vskip 8mm                    % 930715: 8mm white space above title
    \raggedright                  % Front-matter text is ragged right
    \hyphenpenalty\@M             % and is not hyphenated.
    {\Large \@title \par}         % Title set in larger font. 
    \vskip\@bls                   % One line of vertical space after title.
    {\normalsize                  % each author set in the normal 
     \@author \par}               % typeface size 
    \vskip\@bls                   % One line of vertical space after author(s).
    \@address                     % all addresses
  \egroup
  \twocolumn[%                    % Front-matter text is over 2 columns.
    \unvbox\fm@box                % Unwrap contents of front-matter box
    \vskip\@bls                   % add 1 line of vertical space, 
    \unvbox\abstract@box          % unwrap contents of abstract boxes,
    \vskip 2pc]}                  % and add 2pc of vertical space

\newcounter{address} 
\def\theaddress{\alph{address}}
\def\@makeadmark#1{\hbox{$^{\rm #1}$}}   

\def\address#1{\addressmark\begingroup
  \xdef\@tempa{\theaddress}\let\\=\relax
  \def\protect{\noexpand\protect\noexpand}\xdef\@address{\@address
  \protect\addresstext{\@tempa}{#1}}\endgroup}
\def\@address{}

\def\addressmark{\stepcounter{address}%
  \xdef\@tempb{\theaddress}\@makeadmark{\@tempb}}

\def\addresstext#1#2{\leavevmode \begingroup
  \raggedright \hyphenpenalty\@M \@makeadmark{#1}#2\par \endgroup
  \vskip\@bls}

\newbox\abstract@box              % Box to capture abstract in

\def\abstract{%
  \global\setbox\abstract@box=\vbox\bgroup
  \small\rm
  \ignorespaces}
\def\endabstract{\par \egroup}

\def\thebibliography#1{\section*{REFERENCES}\list{\arabic{enumi}.}
  {\settowidth\labelwidth{#1.}\leftmargin=1.67em
   \labelsep\leftmargin \advance\labelsep-\labelwidth
   \itemsep\z@ \parsep\z@
   \usecounter{enumi}}\def\makelabel##1{\rlap{##1}\hss}%
   \def\newblock{\hskip 0.11em plus 0.33em minus -0.07em}
   \sloppy \clubpenalty=4000 \widowpenalty=4000 \sfcode`\.=1000\relax}

\newcount\@tempcntc
\def\@citex[#1]#2{\if@filesw\immediate\write\@auxout{\string\citation{#2}}\fi
  \@tempcnta\z@\@tempcntb\m@ne\def\@citea{}\@cite{\@for\@citeb:=#2\do
    {\@ifundefined
       {b@\@citeb}{\@citeo\@tempcntb\m@ne\@citea
        \def\@citea{,\penalty\@m\ }{\bf ?}\@warning
       {Citation `\@citeb' on page \thepage \space undefined}}%
    {\setbox\z@\hbox{\global\@tempcntc0\csname b@\@citeb\endcsname\relax}%
     \ifnum\@tempcntc=\z@ \@citeo\@tempcntb\m@ne
       \@citea\def\@citea{,\penalty\@m}
       \hbox{\csname b@\@citeb\endcsname}%
     \else
      \advance\@tempcntb\@ne
      \ifnum\@tempcntb=\@tempcntc
      \else\advance\@tempcntb\m@ne\@citeo
      \@tempcnta\@tempcntc\@tempcntb\@tempcntc\fi\fi}}\@citeo}{#1}}

\def\@citeo{\ifnum\@tempcnta>\@tempcntb\else\@citea
  \def\@citea{,\penalty\@m}%
  \ifnum\@tempcnta=\@tempcntb\the\@tempcnta\else
   {\advance\@tempcnta\@ne\ifnum\@tempcnta=\@tempcntb \else
\def\@citea{--}\fi
    \advance\@tempcnta\m@ne\the\@tempcnta\@citea\the\@tempcntb}\fi\fi}

\def\ps@crcplain{\let\@mkboth\@gobbletwo
     \def\@oddhead{\reset@font{\sl\rightmark}\hfil \rm\thepage}%
     \def\@evenhead{\reset@font\rm \thepage\hfil\sl\leftmark}%
     \let\@oddfoot\@empty
     \let\@evenfoot\@oddfoot}

\ps@crcplain                    % modified 'plain' page style

\newcommand{\AmS}{{\protect\the\textfont2
  A\kern-.1667em\lower.5ex\hbox{M}\kern-.125emS}}

\makeatother

\newcommand{\bee}{\begin{equation}}
\newcommand{\ene}{\end{equation}}

\title{The Multiplicative Anomaly of Regularized Functional Determinants}

\author{{Sergio Zerbini}
\address{INFN, Gruppo Collegato di Trento, Sezione di Padova, Italy}
\address{Dipartimento di Fisica, Universit\`a 
degli Studi di Trento, Italy}%
\thanks{Contribution to {\it Quantum Gravity and Spectral geometry}, 
Napoli July 1-7, 2001.}}

\begin{document}
\begin{abstract}
The multiplicative anomaly related to the functional regularized determinants
involving products of elliptic operators is introduced and some of its 
properties discussed. Its relevance concerning 
the mathematical consistency is stressed. With regard to  
 its possible physical relevance, some  examples  are illustrated.

\end{abstract}

\maketitle
It is well known that the vacuum energy related to relativistic quantum
fields defined on topological non-trivial manifolds may lead to interesting 
physical phenomena such as Casimir effects. Within this context, it is 
crucial the relativistic nature of quantum  fields, namely the 
fact that an infinite number of degrees of freedom is involved.  

In the one-loop approximation or in the external field approximation, one may
describe  quantum (scalar) field by means of path (Euclidean) integral and 
expressing the Euclidean partition function as 

\begin{equation}
Z=\left( \det A \right)^{-1/2}
\:,\label{0}
\end{equation}
with $A$ an elliptic self-adjoint non negative differential operator. The 
latter quantity is 
ill defined, since naively, the determinant of a  
self-adjoint elliptic operator $A$, as 
 product of its
eigenvalues, is formally divergent. A large class of regularizations of this 
functional determinant are at disposal and within this class, the finite 
part, modulo a term depending on the renormalization scale, can be evaluated 
by means of the zeta-function regularization. 
When the ($D$-dimensional) manifold is smooth and compact,
the spectrum is discrete and the zeta-function regularized determinant is 
given by \cite{dowker,hawking} (see also \cite{zeta})
$$
\ln\det (A/M^2)
=-\zeta'(0|A)-\zeta(0|A)\ln M^2\,,
$$
where $M^2$ is a renormalization scale mass and $\zeta(s|A)$ is the zeta 
function related to $A$,
an elliptic  operator, analytically continued in the whole complex plane. 
It is possible to show that $s=0$ is regular. Thus 
$\zeta'(0|A)$  exists and gives the value of the regularized functional 
determinant. 

Now one arrives at a technical
 but crucial point: the zeta-function
regularized determinants do {\it not} satisfy the relation 
$$\ln \det
(AB)=\ln \det A+\ln \det B\,.
$$
In fact, in
general,  there exists the so-called multiplicative anomaly, defined as:
$$
a(A,B)=\ln \det (AB)-\ln \det (A)-\ln \det (B)\,,  
$$
with the determinants of the two elliptic operators, $A$ and $B$,
 defined (e.g.,
regularized) by means of the zeta-function. 
This anomaly has been discovered by Wodzicki \cite{wod}. In the simple but 
important case in which $A$ and $B$ are commuting invertible self-adjoint 
elliptic 
operators of second order, the multiplicative anomaly can be directly 
evaluated by the Wodzicki formula
\begin{equation}
a(A,B)
=\frac{1}{8} \, \mbox{res}\left[ (\ln(A B^{-1}))^2 \right]
\:,\label{wod}
\end{equation}
where the non-commutative residue \cite{wod} related to a 
classical pseudo-differential operator $Q$ of order zero may defined by 
the logarithmic term in $t$ of the following heat-kernel expansion
$$
\mbox{Tr}(Qe^{-t A})= \sum_j c_j t^{(j-D)/2}-
\frac{\mbox{res}\:Q}{2} \ln t+O(t \ln t)\,,
$$
where $A$ is an elliptic non negative operator of second order. It is possible to show that $ \mbox{res}\:Q$ doesn not depend on $A$.
The non-commutative residue can also be evaluated by means of
$$
\mbox{res}\, Q =(2\pi)^{-D}\int_{M_D}dx \, \int_{|k|=1}Q_{-D}(x,k)dk\,.
$$
Here the homogeneity component of order $-D$ of the complete symbol appears.
Recall that a classical  pseudo-differential operator $Q$ of order zero 
has a complete 
symbol $ e^{ikx}Qe^{-ikx}$ admitting the following
asymptotics expansion, valid for large $|k|$ 
$$
Q(x,k)\simeq \sum_{j=0}^\infty Q_{-j}(x,k)
$$
where the expansion coefficients satify  the homogeneity property 
$Q_j(x, \lambda k)=\lambda^{-j}Q_{-j}(x,k)$.
Evaluation of the multiplicative anomaly in 
non trivial topological manifold can be found in \cite{will}.

Physical examples where multiplicative anomaly appears are the ones related to 
the presence of internal symmetry with   vector valued fields \cite{eli98}.
For its physical relevance, we shall considered first  a gas of free charged 
Bose fields at finite temperature and non vanishing chemical potential.
We also  assume that the Coulomb interaction can be neglected.  

It is well known that, in this case, the  grand canonical partition function 
reads
\begin{equation}
Z_{\beta,\mu}=\int_{\phi(\tau)=\phi(\tau+\beta)}D\phi_i
e^{-\frac{1}{2}\int_0^\beta d\tau \int d^3x \phi_iA_{ij}\phi_j}
\:,\label{4}
\end{equation}
with 
$$A_{ij}=\left( L_\tau+ L_3-\mu^2
\right)\delta_{ij}+ 2\mu \epsilon_{ij}
\sqrt{L_\tau}\:,
$$
$$
L_3=-\Delta_3+m^2\,,
$$
$\Delta_3$ being the Laplace operator on $R^3$ (continuous
spectrum $\vec k^2$) and
$L_\tau=-\partial^2_\tau$ (discrete spectrum over the Matsubara frequencies
$\omega_n^2=\frac{4\pi^2}{\beta^2}$ ), $\beta$ is the inverse of the 
temperature and $\mu$ is the chemical potential. 
Thus,  the  grand canonical partition function may be written as 
(see, for example, \cite{eli1} and references therein)
\begin{equation}
\ln Z_{\beta,\mu}=-\ln\det \left\| A_{ik}\right\|\,.
\label{o2}\,.
\end{equation}
Now the algebraic determinant, denoted by $|A|$, can be evaluated and gives 
\begin{equation}
| A_{ik}| =( K_+K_-)
\label{o3}\,,
\end{equation}
with
$
K_\pm=L_3+( \sqrt L_\tau \pm i\mu)^2 $.
However, it is easy to show that another factorization exists, i.e. 
\begin{equation}
| A_{ik}| =( L_+L_-)
\label{o4}\,,
\end{equation}
with
$L_\pm=L_\tau+( \sqrt L_3 \pm \mu)^2$.
Of course, one has $| A_{ik}| = L_+L_-=K_+K_-$, and in both 
cases one is dealing with  the product of two pseudo-differential
operators ($\Psi$DOs), the  couple $L_+$ and $L_-$ being also
formally self-adjoint. As a consequence, one has to deal with the product of
two different pair of operators. 
Thus, the partition function may be written as
\begin{equation}
\ln Z_{\beta,\mu} =-\ln\det  K_+ -\ln\det K_-+a(K_+,K_-)\,,
\label{o4}
\end{equation}
but also as
\begin{equation}
\ln Z_{\beta,\mu}
=-\ln\det  L_+ -\ln\det L_-+a(L_+,L_-)\,.
\label{o5}
\end{equation}
The evaluation of the multiplicative anomalies which appear in the above 
expressions can be  done making use of the 
Wodzicki formula and a complete agreement is found between the two expressions
 of the partition function \cite{eli1,toms,evans}.
Thus, if  one {\it neglects} the multiplicative anomaly, one arrives at a 
mathematical inconsistency.

Let us come to the possibile physical relevance. In the above example, 
the multiplicative anomaly 
is linear in $\beta$. As a result, it modifies {\it only} the vacuum 
structure. 
In normal 
situations, namely in absence of spontaneus symmetry breaking, a one-loop 
calculation  shows that it can be re-absorbed  in the one-loop 
renormalization 
process, or in other words, there is no contribution to one-loop beta 
function coming from the multiplicative anomaly \cite{eli98,eli2}. Briefly, 
the argument
goes as follows: the one loop beta-function satifies the one-loop 
renormalization group equations
\begin{equation}
M\frac{d S(\lambda(M))}{dM}=-\frac{1}{2}\zeta(0|A)+\mbox{higher loops}\,,
\end{equation}
where $S$ is the classical action thought as a function of the running 
coupling constant 
$\lambda(M)$ and on the
background field.
Since 
$$A=\mbox{diag}(-\Delta+M_1^2,-\Delta+M_2^2)=-\Delta\, 1 +V\,,
$$ where 
$M_1$ and $M_2$ are effective masses depending on the running coupling 
constant and background field and $1$ 
represents the identity operator, one has
$$
\zeta(0|A)=\frac{1}{2(4 \pi)^2}\int d^4x\: \mbox{tr}\,V^2
$$
$$
=\frac{1}{2(4 \pi)^2}\int d^4x (M_1^4+M_2^4)\,.
$$
Thus, the multiplicative anomaly does not give contribution at one-loop level.

However the situation may change when there exists a spontaneous symmetry 
breaking. In fact, one may perform a charge renormalization such that in the 
symmetry breaking phase, one obtains the usual condition for the critical 
temperature,
with a corrected non relativistic limit. However, in the symmetric phase, 
namely when the temperature is higher than the critical temperature, such 
charge renormalization leaves a sub-leading contribution in the free energy
which depends on the presence of the multiplicative anomaly \cite{eli1}. 

Another interesting and related case has been recently considered 
\cite{filippi} and involves a self-interacting charged Bose field at finite 
temperature. 
This system is highly non trivial, but it can be treated in the large N limit,
in order to deal with higher loop contributions. As a consequence, 
in presence of 
symmetry breaking, standard renormalization can be done, but it seems that  
the presence of the multiplicative anomaly cannot be renormalized away and  
gives rise to additional terms which modify the transition temperature and 
leads to additional new terms for the pressure. With regards to the 
possible physical relevance of the multiplicative anomaly, different 
conclusions  are reported in \cite{toms,evans}.

We conclude with some remarks about first order differential operators. As an
example, we consider the
Dirac  operator $K_0$.  One again should give a meaning to the formal
effective action $\Gamma=\ln \det K_0\,$. Let us start with massless Euclidean 
self-adjoint Dirac operator  $K_0^+=K_0$.
In this case the spectrum of $K_0$
on a compact (curved) manifold $M$ of dimension $D$, is unbounded over the  
whole real axis. If fact $
K_0=i\gamma^\mu\nabla_\mu\:,
\gamma_\mu=e_k^\mu\gamma^k  $
 $e_k^\mu$ are  ``viel-bein'' fields,
$\nabla_\mu$ is the covariant derivative. There exists an   
ambiguity  associated with negative eingenvalues  and there are 
{\it two} inequivalent zeta-functions (see, for example 
\cite{deser,cognola99} and 
references quoted therein).
As a result, one has 
\begin{equation}
\zeta_\pm(s|K_0)=\frac{1}{2}( 1+e^{\mp i\pi s})[ \zeta(\frac{s}{2}|L)
+ \eta(s|K_0)]
\end{equation}
in which $L=K_0^2$ is the spinor Laplacian and  
\begin{equation}
\eta(s|K_0)=
\sum_i \lambda_i^{-s}-\sum_i (-\mu_i)^{-s}
\end{equation}
is the eta function. However, in the 
even dimensional case  $D=2p$, 
$\gamma^{D+1}=\gamma^1\cdots\gamma^D$ exists, and eta-function is vanishing 
by symmetry. As a result, the effective action is proportional to 
\begin{equation}
\zeta'_{\pm}(0|K_0)= \frac{1}{2}\zeta'(0|L) \mp \frac{i \pi}{2} A_p(L)
\end{equation}
Here $A_p(L)$ are the Seeley-De Witt integral coefficients associated with the 
second order elliptic operator $L$.
In this case, the ambiguity depends on the local functionals  $A_p(L)$, but 
such terms may be reassorbed by the renormalization process.

In the odd dimensional case  $D=2p+1$, there is no symmetry in the Dirac 
spectrum and the spinor effective action reads 
\begin{equation}
\zeta'_{\pm}(0|K_0)= \frac{1}{2}\zeta'(0|L) \pm \frac{i \pi}{2} \eta(0|K_0)
\label{bbb}\,.
\end{equation}
Here, the eta invariant $\eta(0|A)$ is a {\it non} local functional of the 
external geometrical fields and it {\it cannot}  
be removed by the renormalization.  

The multiplicative anomaly appears in the massive case,  when the Euclidean 
Dirac  operator$K=K_0+im$ is not hermitian, the eingenvalues being 
$\lambda_i+im$. In this case, it is possible to show that 
there is no ambiguity, but again the odd dimensional case is {\it non } 
trivial and we have \cite{cognola99}

$$
\zeta_{+}'(0|K)=\frac{1}{2}\:\zeta'(0|L+m^2)+  
\frac{i\pi}{2}\eta(0|K_0)\nonumber 
$$
$$
-is(K_0,m)
-\frac{\pi}{2}\sum_{j=0}^{p}\frac{(-1)^j 
m^{2j+1}A_{p-j}(L) }{\Gamma(j+\frac{3}{2})}\,
$$
where the eta-like function contribution appears. It can be evaluated  by 
means of
$$
s(K_0,m)
=\int_0^\infty \frac{\sin m t}{t}
\mbox{Tr}( \frac{K_o}{|K_0|} e^{-t|K_0|}) \nonumber
$$
As a result, 
the Folk Theorem $\det K=\det K^+=\sqrt{\det K^+K}$ strictly speaking, 
{\it does not} hold. In fact, the correct formulae for the relevant 
functional spinor determinats are, for $D$ even
$$
\frac{\det K}{\sqrt{\det (L+m^2)}}=e ^{- i \frac{\pi}{2} \zeta(0|L+m^2)
-\frac{1}{2} a_{2p}(K^+,K)} \nonumber\,
$$
and  for $D$ odd
$$
\frac{\det K}{\sqrt{\det (L+m^2)}} = e^{-i \frac{\pi}{2} \eta(0|A)+i 
s(A,m) -\frac{1}{2}a_{2p+1}(K^+,K)}
\,,
$$
where the multiplicative anomalies are given respectively by
$$
a_{2p}(K^+,K)=2\sum_{j=1}^{p}\frac{(-1)^j m^{2j}c_j}{j!}A_{p-j}(L)\:
\nonumber\,,
$$
with
$$
c_j=\sum_{l=1}^j\frac{1}{2l-1}
$$
and
$$
a_{2p+1}(K^+,K)=\pi\sum_{j=0}^{p}
\frac{(-1)^{j+1}m^{2j+1}}{\Gamma(j+\frac{3}{2})}\:A_{p-j}(L)\nonumber
$$
As an example, for $D=4$, one has
$$
a_4(K^+,K)=\frac{4}{3}A_0(L)m^4-2m^2A_1(L).
$$
It should be noted that the multiplicative anomaly term induces 
a Einstein-Hilbert term, linear in the scalar curvature, contained in the 
coefficient $A_1(L)$. 

Finally, we also would like mention two other issues where the multiplicative
anomaly seems to play a role. The first one is the relationship beteewen the 
so 
called covariant and consistent anomaly associated with vector-axial vector
non abelian gauge symmetry. In two dimensions, it has been proved in 
\cite{cogn} that the difference between the consistent and covariant anomaly
can be thought as a multiplicative anomaly effect. The second issue is 
related to the so called dimensional  reduction anomaly \cite{frolov,co},
an anomaly which is present in quantum field theory once a  
dimensional reduction is performed and a truncation of the related  harmonic 
sum is done.

\end{document}